\documentclass[aps,prl,reprint,superscriptaddress,showpacs]{revtex4-1}
\usepackage{graphicx}

\begin{document}


\title{Reduction of thermal fluctuations in a cryogenic laser interferometric gravitational wave detector}


\author{Takashi Uchiyama}
\email[E-mail me at: ]{uchiyama@icrr.u-tokyo.ac.jp}
\altaffiliation{}
\affiliation{Kamioka Observatory, Institute for Cosmic Ray Research, the University of Tokyo, 456 Higashi-Mozumi, Kamioka, Hida Gifu 506-1205, Japan.}

\author{Shinji Miyoki}
\affiliation{Institute for Cosmic Ray Research, the University of Tokyo, 5-1-5 Kashiwa-no-ha, Kashiwa, Chiba 277-8582, Japan.}

\author{Souichi Telada}
\affiliation{The National Institute of Advanced Industrial Science and Technology, 1-1-1 Umezono, Tsukuba, Ibaraki 305-8568, Japan.}

\author{Kazuhiro Yamamoto}
\altaffiliation[Present address: ]{Institute for Cosmic Ray Research, the University of Tokyo, 5-1-5 Kashiwa-no-ha, Kashiwa, Chiba 277-8582, Japan.}
\affiliation{Max-Planck-Institut f\"{u}r Gravitationsphysik (Albert-Einstein-Institut) and Institut f\"{u}r Gravitationsphysik, Leibniz Universit\"{a}t Hannover, Callinstrasse 38, D-30167 Hannover, Germany.}

\author{Masatake Ohashi}
\affiliation{Institute for Cosmic Ray Research, the University of Tokyo, 5-1-5 Kashiwa-no-ha, Kashiwa, Chiba 277-8582, Japan.}

\author{Kazuhiro Agatsuma}
\altaffiliation[Present address: ]{National Astronomical Observatory of Japan, 2-21-1 Osawa, Mitaka, Tokyo 181-8588, Japan.}
\affiliation{Institute for Cosmic Ray Research, the University of Tokyo, 5-1-5 Kashiwa-no-ha, Kashiwa, Chiba 277-8582, Japan.}

\author{Koji Arai}
\altaffiliation[Present address: ]{LIGO project, California Institute of Technology 100-36, Pasadena, CA 91125, USA. }
\affiliation{National Astronomical Observatory of Japan, 2-21-1 Osawa, Mitaka, Tokyo 181-8588, Japan.}

\author{Masa-Katsu Fujimoto}
\affiliation{National Astronomical Observatory of Japan, 2-21-1 Osawa, Mitaka, Tokyo 181-8588, Japan.}

\author{Tomiyoshi Haruyama}
\affiliation{High Energy Accelerator Research Organization, KEK, 1-1 Oho, Tsukuba, Ibaraki 305-0801, Japan.}

\author{Seiji Kawamura}
\altaffiliation[Present address: ]{Institute for Cosmic Ray Research, the University of Tokyo, 5-1-5 Kashiwa-no-ha, Kashiwa, Chiba 277-8582, Japan.}
\affiliation{National Astronomical Observatory of Japan, 2-21-1 Osawa, Mitaka, Tokyo 181-8588, Japan.} 


\author{Osamu Miyakawa}
\affiliation{Kamioka Observatory, Institute for Cosmic Ray Research, the University of Tokyo, 456 Higashi-Mozumi, Kamioka, Hida Gifu 506-1205, Japan.}

\author{Naoko Ohishi}
\altaffiliation[Present address: ]{National Astronomical Observatory of Japan, 2-21-1 Osawa, Mitaka, Tokyo 181-8588, Japan.}
\affiliation{Kamioka Observatory, Institute for Cosmic Ray Research, the University of Tokyo, 456 Higashi-Mozumi, Kamioka, Hida Gifu 506-1205, Japan.}

\author{Takanori Saito}
\affiliation{Institute for Cosmic Ray Research, the University of Tokyo, 5-1-5 Kashiwa-no-ha, Kashiwa, Chiba 277-8582, Japan.}

\author{Takakazu Shintomi}
\affiliation{Advanced Research Institute for the Science and Humanities, Nihon University, 12-5 Goban-cho, Chiyoda-ku, Tokyo 102-8251, Japan.}

\author{Toshikazu Suzuki}
\affiliation{High Energy Accelerator Research Organization, KEK, 1-1 Oho, Tsukuba, Ibaraki 305-0801, Japan.}

\author{Ryutaro Takahashi}
\altaffiliation[Present address: ]{Institute for Cosmic Ray Research, the University of Tokyo, 5-1-5 Kashiwa-no-ha, Kashiwa, Chiba 277-8582, Japan.}
\affiliation{National Astronomical Observatory of Japan, 2-21-1 Osawa, Mitaka, Tokyo 181-8588, Japan.} 

\author{Daisuke Tatsumi}
\affiliation{National Astronomical Observatory of Japan, 2-21-1 Osawa, Mitaka, Tokyo 181-8588, Japan.}


\date{\today}

\begin{abstract}
The thermal fluctuation of mirror surfaces is the fundamental limitation for interferometric gravitational wave (GW) detectors. Here, we experimentally demonstrate for the first time a reduction in a mirror's thermal fluctuation in a GW detector with sapphire mirrors from the Cryogenic Laser Interferometer Observatory at 17\,K and 18\,K. The detector sensitivity, which was limited by the mirror's thermal fluctuation at room temperature, was improved in the frequency range of 90\,Hz to 240\,Hz by cooling the mirrors. The improved sensitivity reached a maximum of $2.2 \times 10^{-19}\,\textrm{m}/\sqrt{\textrm{Hz}}$ at 165\,Hz.





\end{abstract}


\maketitle

\paragraph{Introduction}

Two hundred years ago, Robert Brown investigated the random motion of small particles in water \cite{brownian}. This random motion is now understood to be an irreducible natural phenomenon, and it creates a fundamental limit on the precision of measurements, including the measurement of fundamental constants, high-resolution spectroscopy and fundamental physics experiments using a frequency-stabilized laser \cite{freq_standard} and gravitational wave (GW) detection by a laser interferometer with suspended mirrors \cite{salson_prd}.





In the case of interferometric GW detectors, thermal noise in the mirror typically limits the detector sensitivity to approximately a few hundred Hz, which lies in the important frequency region for the detection of GWs from the binary coalescence of neutron stars. Although kilometer-scale first-generation laser interferometric GW detectors, such as Laser Interferometer Gravitational-Wave Observatory (LIGO) \cite{ligo} and VIRGO \cite{virgo}, have already performed several long-term observation runs, no GW signal has yet been observed. The sensitivity must be improved by one order of magnitude to be able to detect GWs within a single year of observation. 

According to the fluctuation-dissipation theorem \cite{fdt}, the power of thermal fluctuations is proportional to both temperature and mechanical loss. Thus, a mirror constructed of a low-loss material operating at a cryogenic temperature (a cryogenic mirror) is a good candidate for creating a low-thermal-fluctuation mirror. Because of various difficulties associated with implementing cryogenic mirrors in interferometers, major efforts have thus far been devoted to finding low-loss materials at room temperature.

We originally proposed using a suspended sapphire mirror cooled to less than 20\,K for a GW detector \cite{sapphire_cooling}, and we have researched and developed this technique \cite{sapphire_subQ, sapphire_fiberQ, coating_Q, sapphire_thermalCond, sapphire_optabsorb, cryogenic_contami, clik_lism, thermal_lens}. Here, we demonstrate for the first time a reduction in mirror thermal fluctuations using this cryogenic mirror in a working GW detector. This reduction is the primary purpose of the Cryogenic Laser Interferometer Observatory (CLIO) \cite{clio}, which is the first GW detector to use a cryogenic mirror.


\paragraph{Experiment}
CLIO was built at an underground site in the Kamioka mine, which is located 220 km northwest of Tokyo in Japan. CLIO is a Michelson interferometer with 100-m Fabry-Perot (FP) arm cavities, each consisting of front and end mirrors \cite{clio, yamamoto_clio}. The front mirrors are suspended closest to the beam splitter and were cooled to 17\,K and 18\,K for the sensitivity measurement reported in this study. The end mirrors remained at 299\,K.

The cavity mirror substrate material is sapphire. The cylindrical substrate has a diameter of 100\,mm, a thickness of 60\,mm and a mass of 1.8\,kg. One surface of the substrate has a highly reflective multilayered film coating of SiO$_{2}$ and Ta$_{2}$O$_{5}$. The mirror is suspended at the final stage of a six-stage suspension system \cite{suspension_clio}, and the length of the wire suspending the mirror is 400\,mm. The suspension system is installed in a cryostat with two layers of radiation shielding (an outer and an inner shield) \cite{suspension_clio}. A two-stage pulse-tube cryo-cooler \cite{cryocooler} cools the outer and inner shields to approximately 70\,K and 10\,K, respectively. It took approximately 250 hours to cool the mirrors, and the vacuum pressure was less than $10^{-4}$\,Pa.

To cool the front mirrors, we changed the mirror suspension wires from Bolfur to 99.999\% purity aluminum to provide higher thermal conductivity, and we added three heat links to each suspension system \cite{yamamoto_clio}. Bolfur wire is an amorphous metal wire made by Unitika, Ltd. with a diameter of 50\,$\mu$m. The aluminum wire used has a diameter of 0.5\,mm. The same aluminum wire was used for the heat links, which provide thermal conduction between the suspended masses (Damping Stage, Cryo-base and Upper Mass) and the inner shield in the cryostat \cite{suspension_clio}. The lengths of the heat links between the inner shield and the Cryo-base, between the Cryo-base and the Upper Mass and between the Damping Stage and the inner shield were 315\,mm, 115\,mm and\,150 mm, respectively; each had one heat link. Thermometers were attached to the Cryo-base and the Upper Mass. We estimated the temperature of the mirror from the temperatures of the Cryo-base and the Upper Mass and from the thermal conductivity of the mirror suspension wires and the heat link. 


One 100-m FP cavity serves as a reference for the laser frequency stabilization, and the length of the other 100-m FP cavity is controlled to maintain the optical resonance \cite{clio}. Coil-magnet actuators, consisting of magnets glued to the mirror and coils facing toward the magnets, are used to control the cavity length. The GW signal is included in the feedback signal for this length control.  In this study, the sensitivity was characterized by the power spectrum density of the displacement in units of m/$\sqrt{\textrm{Hz}}$ and is calculated using the following three measurements: the feedback signal, V/$\sqrt{\textrm{Hz}}$, the loop gain of the length control system, and the response function of the coil-magnet actuators, m/V.

The thermal fluctuation of a mirror surface is caused by several different loss mechanisms. The thermoelastic damping and the internal frictional loss of the sapphire substrate and of the reflective coating films were considered in this study. In the case of the sapphire mirror at room temperature, the largest loss mechanism is the thermoelastic damping of the sapphire substrate (thermoelastic noise). At temperature below 20\,K, the internal frictional loss of the reflective coating films is expected to be the largest loss mechanism. 

The theory of thermoelastic noise \cite{braginsky, cerdonio} has previously been validated experimentally \cite{black}. The power spectrum density ($\textrm{m}^2/\textrm{Hz}$) of the thermoelastic noise in a mirror was shown in Black et al. \cite{black} and described as follows:
\begin{equation}
S_{\mathrm{TE}}(\omega) = \frac{4}{\sqrt{\pi}}\frac{\alpha^2(1+\sigma)^2}{\kappa}k_{\mathrm{B}}T^2WJ(\Omega),
\end{equation}
where $J(\Omega)$ is
\begin{equation}
J(\Omega)=\frac{\sqrt{2}}{\pi ^{3/2}} \int_{0}^{\infty}du \int_{-\infty}^{+\infty}dv\frac{u^3e^{-u^2/2}}{(u^2+v^2)[(u^2+v^2)^2+\Omega^2]},
\end{equation}
and $\Omega$ is 
\begin{equation}
\Omega = \frac{\omega}{\omega_{\mathrm{c}}},
\end{equation}
where $\omega_{\mathrm{c}}$ is 
\begin{equation}
\omega_{\mathrm{c}}=\frac{2\kappa}{\rho CW^2}.
\end{equation}
In these equations, $\omega$ is the angular frequency, $\alpha$ is the thermal expansion coefficient, $\sigma$ is the Poisson's ratio, $\kappa$ is the thermal conductivity, $\rho C$ is the specific heat per unit volume, $k_\mathrm{B}$ is the Boltzmann constant, $T$ is the temperature of the mirror, and $W$ is the beam spot radius on the mirror. In the case of the CLIO mirror at room temperature, $\Omega \gg 1$ is satisfied near 100\,Hz. In this case, the power spectrum density ($\textrm{m}^2/\textrm{Hz}$) is simplified as follows:
\begin{equation}
S_{\mathrm{TE}}^{\Omega \gg 1}(\omega) = \frac{16}{\sqrt{\pi}}\alpha^2(1+\sigma)^2\frac{k_{\mathrm{B}}T^2\kappa}{(\omega \rho C)^2}\frac{1}{W^3}.
\end{equation}
The power spectrum density ($\textrm{m}^2/\textrm{Hz}$) of the thermal noise in a mirror caused by the internal frictional loss of the substrate and of the coating films was shown in Nakagawa et al. \cite{nakagawa} as follows:
\begin{equation}
S(f) = \frac{2k_BT(1-\sigma^2)}{\pi^{3/2}fWE}\phi_{\mathrm{substr}}\{1+\frac{2}{\sqrt{\pi}}\frac{(1-2\sigma)}{(1-\sigma)}\frac{\phi_{\mathrm{coat}}}{\phi_{\mathrm{substr}}}(\frac{d}{W})\}.
\end{equation}
In this equation, $E$ is the Young's modulus, $d$ is the thickness of the coating films and $\phi_{\mathrm{substr}}$ and $\phi_{\mathrm{coat}}$ are the internal frictional loss of the substrate and of the coating films, respectively.





\paragraph{Results}

 \begin{figure}
 \includegraphics[width=0.85\columnwidth]{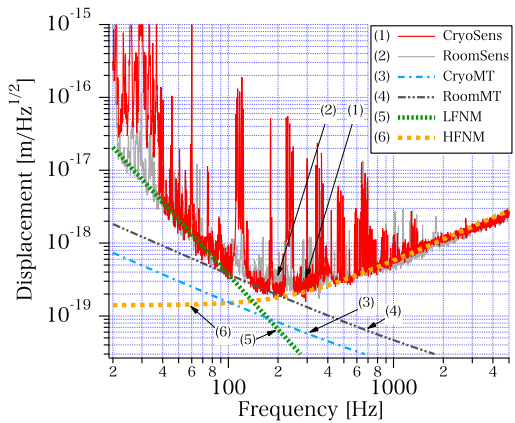}
 \caption{\label{fig1}Comparison of CLIO displacement sensitivity curves. Cryogenic sensitivity (CryoSens) was measured with the front mirrors at 18\,K and 17\,K and with the end mirrors at room temperature (299\,K). Room-temperature sensitivity (RoomSens) was measured with all of the mirrors at 299\,K. CryoSens and RoomSens were measured on March 20, 2010 and November 5, 2008, respectively. The magnified  sensitivity curves are shown in Fig. \ref{fig2}. Mirror thermal noise estimate curves corresponding to each of the sensitivity measurements are also shown (CryoMT and RoomMT). The CryoMT and the RoomMT were estimated to be $1.5\times10^{-19}\,\textrm{m}/ \sqrt{\textrm{Hz}}\times(100\,\textrm{Hz}/f )$ and $3.7\times10^{-19}\,\textrm{m}/\sqrt{\textrm{Hz}}\times(100\,\textrm{Hz}/f )$, respectively. The low-frequency noise model (LFNM) consists of the fitted lines for the noise floor level of CryoSens from 40\,Hz to 70\,Hz and for RoomSens from 20\,Hz to 70\,Hz. The LFNM was estimated to be $3.7\times 10^{-19}\,\textrm{m}/\sqrt{\textrm{Hz}}\times (100\,\textrm{Hz}/f)^{2.5}$. The high-frequency noise model (HFNM) consists of the fitted lines for the noise floor levels of both sensitivity curves above 400\,Hz. Based on the characteristics of the 100-m Fabry-Perot cavities of CLIO, the HFNM was found to be $1.4\times10^{-19}\,\textrm{m}/\sqrt{\textrm{Hz}}\times \sqrt{1+(f/250\,\textrm{Hz})^2}$.}
 \vspace{-15pt}
 \end{figure}

For comparison, Fig. \ref{fig1} presents both the displacement sensitivity curve measured with the front CLIO mirrors cooled to 17\,K and 18\,K (cryogenic sensitivity; CryoSens) as well as the curve without cooled mirrors (room temperature sensitivity; RoomSens). CryoSens and RoomSens were measured on March 20, 2010 and November 5, 2008, respectively. The loop gain of the length control was measured immediately after each feedback signal measurement, and the response function of the coil-magnet actuators was calibrated for each experimental configuration. A sensitivity curve consists of frequency-dependent noise floors and multiple line noises. The noise floor level of CryoSens from 90\,Hz to 240\,Hz is below the noise floor level of RoomSens. By reducing this noise floor level, the detection range for GW signals from the binary coalescence of neutron stars in the optimal direction was improved from 150\,kpc to 160\,kpc. The noise floor at 165\,Hz was reduced from $3.1\times10^{-19}\,\textrm{m}/\sqrt{\textrm{Hz}}$ to $2.2\times10^{-19}\,\textrm{m}/\sqrt{\textrm{Hz}}$  by cooling the front mirrors.

\paragraph{Discussion}

The noise floors of the two sensitivities shown in Fig. \ref{fig1} are very similar from 40\,Hz to\,90 Hz and from 400\,Hz to 5\,kHz, although additional line noises appeared in CryoSens. The line noise near 30\,Hz results from the mechanical resonance of the cooled front mirror suspension systems. The line noise near 120\,Hz and its higher-order harmonics are also caused by mechanical resonance in the suspension wires of the cooled front mirrors. The line noise at 60\,Hz and its higher harmonics are similar to electric power line noise.

 \begin{table*} 
 \caption{\label{table:parameters}Parameters for sapphire mirror thermal noise estimation}
 \begin{ruledtabular}
 \begin{tabular}{clccc}
& Objects &Unit & Room temperature & Cryogenic  \\ \hline
& Mirror temperature & K & $299$ & $17\, \textrm{and}\, 18$ \\  \hline
$W _{\mathrm{front}}$ &Beam spot radius on front mirrors & mm & $4.9$ & $4.9$\\
$W _{\mathrm{end}}$ &Beam spot radius on end mirrors & mm & $8.5$ & $8.5$\\ \hline
& Material properties of Sapphire & & & \\ 
$\alpha$ &Thermal expansion coefficient \cite{data_book, th_expansion_cryo} & 1/K& $5.4 \times 10^{-6}$ & $5.6 \times 10^{-9}$\\
$\rho C$ &Specific heat per unit volume \cite{data_book} & J/K/m$^{3}$& $3.1 \times 10^{6}$ & $2.8 \times 10^{3}$\\
$\kappa$ &Thermal conductivity \cite{data_book} & W/m/K& $46$ & $1.6 \times 10^{4}$\\
$\sigma$ &Poisson's ratio\footnote{We found various values for the Poisson's ratio of sapphire between 0.23 and 0.30. We used the averaged value. 
}& & $0.27$ & $0.27$\\
$E$ &Young's modulus \cite{coating_Q}& Pa& $40 \times 10^{10}$ & $40\times 10^{10}$\\ 
$\phi _{\mathrm{substr}}$ &Mechanical loss \cite{sapphire_subQ}& &  $1/4.6 \times 10^{6}$ & $1/1.5 \times 10^{8}$\\ \hline
$\phi _{\mathrm{coat}}$ &Mechanical loss in coating films \cite{coating_Q} & & $4.0 \times 10^{-4}$ & $4.0 \times 10^{-4}$\\
$d$ &Thickness of coating films & $\mathrm{\mu}$ m & $3.9$ & $3.9$\\
\end{tabular}
 \end{ruledtabular}
 \end{table*}

Figure \ref{fig1} also shows the mirror thermal noise estimates corresponding to each sensitivity measurement (CryoMT and RoomMT), with the fitted lines for the noise floor in low-frequency region (the low-frequency noise model; LFNM) and high-frequency region (the high-frequency noise model; HFNM).

The parameters for the thermal noise calculation are summarized in Table \ref{table:parameters}. We did not perform any fits to the parameters. The RoomMT was calculated to be $3.7\times10^{-19}\,\textrm{m}/\sqrt{\textrm{Hz}}\times(100\,\textrm{Hz}/f )$. Because the mirror thermal noise is larger with a smaller laser beam, the thermal noise of the front mirror is approximately twice as large as the thermal noise of the end mirror in the RoomMT. The CryoMT was greatly reduced to $1.5\times10^{-19}\,\textrm{m}/ \sqrt{\textrm{Hz}}\times(100\,\textrm{Hz}/f )$ by the decrease in the thermal fluctuation of the cooled front mirrors. If all CLIO mirrors are cooled to 20\,K, the mirror thermal noise near 100\,Hz should be reduced to $1.7\times10^{-20}\,\textrm{m}/ \sqrt{\textrm{Hz}}\times(100\,\textrm{Hz}/f )^{1/2}$.

%

The LFNM was estimated from the noise floor of CryoSens between 40\,Hz and 70\,Hz and the noise floor of RoomSens between 20\,Hz and 70\,Hz, with a value of  $3.7\times10^{-19}\,\textrm{m}/\sqrt{\textrm{Hz}}\times(100\,\textrm{Hz}/f )^{2.5}$. A discussion of the origin of the LFNM is outside the scope of the current work. The HFNM was estimated from the noise floors of both sensitivity curves above 400\,Hz by assuming a simple pole frequency dependence \cite{salson_book}. The HFNM consists of photocurrent shot noise and laser intensity noise at a frequency of 15.8\,MHz and was estimated to have a value of $1.4\times10^{-19}\,\textrm{m}/\sqrt{\textrm{Hz}}\times\sqrt{1+(f/250\,\textrm{Hz})^2}$; the HFNM depends on the length and Finesse of the 100-m FP cavity. Cavity pole frequencies of 262 Hz and 246 Hz were obtained for the measurements of CryoSens and RoomSens, respectively. The systematic error in the noise floor amplitude due to the discrepancy in the cavity pole frequencies was less than 2\% at frequency below 200\,Hz.

 \begin{figure}
 \includegraphics[width=0.85\columnwidth]{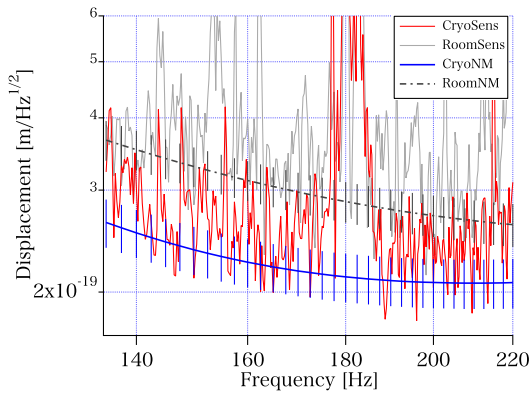}
 \caption{\label{fig2}Comparison of CLIO displacement sensitivity curves and noise models. The cryogenic sensitivity (CryoSens) and the room-temperature sensitivity (RoomSens) show similar curves, as observed in Fig. \ref{fig1}. The cryogenic noise model (CryoNM) is the noise-floor model curve for CryoSens and is calculated as the quadrature sum of the low-frequency noise model (LFNM), the high-frequency noise model (HFNM) and the mirror thermal noise estimate for CryoSens (CryoMT). The room-temperature noise model (RoomNM) is the noise-floor model curve for RoomSens and is calculated as the quadrature sum of the LFNM, the HFNM and the mirror thermal noise estimate for RoomSens (RoomMT). Error bars are shown for both CryoNM and RoomNM, representing the quadrature sums of the calibration error and the noise estimation errors. We estimated a systematic calibration error of $\pm$2.5\%, a $\pm$5\% estimation error for CryoMT and RoomMT and a $\pm$10\% statistical error for LFNM and HFNM. The noise estimation error is the quadrature sum of the error in the mirror thermal noise estimate (CryoMT or RoomMT) and the errors in the LFNM and HFNM.}
 \vspace{-15pt}
 \end{figure}

Figure \ref{fig2} compares CryoSens, RoomSens, the cryogenic noise model (CryoNM) and the room-temperature noise model (RoomNM). CryoNM is the quadrature sum of the LFNM, the HFNM and CryoMT. RoomNM is the quadrature sum of the LFNM, the HFNM and RoomMT. Error bars are shown for both CryoNM and RoomNM, representing the quadrature sum of the calibration error and the noise estimation error. We estimated a $\pm$2.5\% systematic error for the calibration, a $\pm$5\% estimation error for CryoMT and RoomMT and a $\pm$10\% statistical error for LFNM and HFNM. The noise estimation error is the quadrature sum  of the error in the mirror thermal noise estimate (CryoMT or RoomMT) and the errors in LFNM and HFNM. 

The noise floor of RoomSens agrees with RoomNM. The noise floor of CryoSens is below RoomNM and agrees instead with CryoNM. The difference between CryoNM and RoomNM is solely based on the mirror thermal noise estimate. Thus, we conclude that the sensitivity limitation due to thermal noise in the mirror at room temperature was improved by the noise reduction because of the decreased thermal fluctuation in the cooled front mirrors. This observation is the first demonstration that mirrors display less thermal fluctuation at cryogenic temperatures than at room temperature.

\paragraph{Conclusion}
The thermal fluctuation of mirror surfaces represents the fundamental limitation on experiments that require high-precision measurements, such as experiments using a frequency-stabilized laser and GW detection using a laser interferometer. The cryogenic mirror technique uses a low-mechanical-loss mirror at a low temperature to reduce such fluctuations. Our demonstration using an actual GW detector directly proves that the cryogenic mirror technique effectively improves the sensitivity of the GW detector. This cryogenic mirror technique will be used in advanced GW detectors, such as KAGRA formerly called the Large-Scale Cryogenic Gravitational Wave Telescope \cite{lcgt}, on which construction began in the middle of 2010 in Japan, and the Einstein Telescope project in Europe \cite{et}. The mirror thermal noise of KAGRA near 100\,Hz, which limits the sensitivity to $4\times10^{-20}\,\textrm{m}/ \sqrt{\textrm{Hz}}\times(100\,\textrm{Hz}/f )$ at 299\,K, will be reduced to $5\times10^{-21}\,\textrm{m}/ \sqrt{\textrm{Hz}}\times(100\,\textrm{Hz}/f )^{1/2}$ at 20\,K due to the cryogenic mirror technique. The KAGRA sensitivity will make it possible to detect a GW signal from the binary coalescence of neutron stars in the optimal direction up to 250\,Mpc with a signal-to-noise ratio of 10 and the event rate is expected to be approximately 6 events per year \cite{lcgt}. We believe that our achievement represents a breakthrough in the study of thermal fluctuations, laser frequency stabilization and GW detection.

\begin{acknowledgments}
 We wish to thank our CLIO, TAMA and KAGRA collaborators, the Kamioka Observatory, Institute for Cosmic Ray Research, the University of Tokyo and Dr. Hiroaki Yamamoto of the LIGO laboratory for his advice on the manuscript. This work was supported in part by a Grant-in-Aid for Scientific Research on Priority Areas (No. 415) from the Ministry of Education, Culture, Sports, Science and Technology (MEXT). This work was also supported in part by a Grant-in-Aid for Scientific Research (A, No. 18204021) by the Japan Society for the Promotion of Science (JSPS).
\end{acknowledgments}

\bibliography{clio_prl}

\providecommand{\noopsort}[1]{}\providecommand{\singleletter}[1]{#1}%
\begin{thebibliography}{28}%
\makeatletter
\providecommand \@ifxundefined [1]{%
 \@ifx{#1\undefined}
}%
\providecommand \@ifnum [1]{%
 \ifnum #1\expandafter \@firstoftwo
 \else \expandafter \@secondoftwo
 \fi
}%
\providecommand \@ifx [1]{%
 \ifx #1\expandafter \@firstoftwo
 \else \expandafter \@secondoftwo
 \fi
}%
\providecommand \natexlab [1]{#1}%
\providecommand \enquote  [1]{``#1''}%
\providecommand \bibnamefont  [1]{#1}%
\providecommand \bibfnamefont [1]{#1}%
\providecommand \citenamefont [1]{#1}%
\providecommand \href@noop [0]{\@secondoftwo}%
\providecommand \href [0]{\begingroup \@sanitize@url \@href}%
\providecommand \@href[1]{\@@startlink{#1}\@@href}%
\providecommand \@@href[1]{\endgroup#1\@@endlink}%
\providecommand \@sanitize@url [0]{\catcode `\\12\catcode `\$12\catcode
  `\&12\catcode `\#12\catcode `\^12\catcode `\_12\catcode `\%12\relax}%
\providecommand \@@startlink[1]{}%
\providecommand \@@endlink[0]{}%
\providecommand \url  [0]{\begingroup\@sanitize@url \@url }%
\providecommand \@url [1]{\endgroup\@href {#1}{\urlprefix }}%
\providecommand \urlprefix  [0]{URL }%
\providecommand \Eprint [0]{\href }%
\providecommand \doibase [0]{http://dx.doi.org/}%
\providecommand \selectlanguage [0]{\@gobble}%
\providecommand \bibinfo  [0]{\@secondoftwo}%
\providecommand \bibfield  [0]{\@secondoftwo}%
\providecommand \translation [1]{[#1]}%
\providecommand \BibitemOpen [0]{}%
\providecommand \bibitemStop [0]{}%
\providecommand \bibitemNoStop [0]{.\EOS\space}%
\providecommand \EOS [0]{\spacefactor3000\relax}%
\providecommand \BibitemShut  [1]{\csname bibitem#1\endcsname}%
\let\auto@bib@innerbib\@empty
\bibitem [{\citenamefont {Brown}(1828)}]{brownian}%
  \BibitemOpen
  \bibfield  {author} {\bibinfo {author} {\bibfnamefont {R.}~\bibnamefont
  {Brown}},\ }\href@noop {} {\bibfield  {journal} {\bibinfo  {journal} {Phil.
  Mag.}\ }\textbf {\bibinfo {volume} {4}},\ \bibinfo {pages} {161} (\bibinfo
  {year} {1828})}\BibitemShut {NoStop}%
\bibitem [{\citenamefont {Numata}\ \emph {et~al.}(2004)\citenamefont {Numata},
  \citenamefont {Kemery},\ and\ \citenamefont {Camp}}]{freq_standard}%
  \BibitemOpen
  \bibfield  {author} {\bibinfo {author} {\bibfnamefont {K.}~\bibnamefont
  {Numata}}, \bibinfo {author} {\bibfnamefont {A.}~\bibnamefont {Kemery}}, \
  and\ \bibinfo {author} {\bibfnamefont {J.}~\bibnamefont {Camp}},\ }\href@noop
  {} {\bibfield  {journal} {\bibinfo  {journal} {Phys.\ Rev.\ Lett}\ }\textbf
  {\bibinfo {volume} {93}},\ \bibinfo {pages} {250602} (\bibinfo {year}
  {2004})}\BibitemShut {NoStop}%
\bibitem [{\citenamefont {Saulson}(1990)}]{salson_prd}%
  \BibitemOpen
  \bibfield  {author} {\bibinfo {author} {\bibfnamefont {P.~R.}\ \bibnamefont
  {Saulson}},\ }\href@noop {} {\bibfield  {journal} {\bibinfo  {journal}
  {Phys.\ Rev.\ D}\ }\textbf {\bibinfo {volume} {42}},\ \bibinfo {pages} {2437}
  (\bibinfo {year} {1990})}\BibitemShut {NoStop}%
\bibitem [{\citenamefont {Abramovici}\ \emph {et~al.}(1992)\citenamefont
  {Abramovici} \emph {et~al.}}]{ligo}%
  \BibitemOpen
  \bibfield  {author} {\bibinfo {author} {\bibfnamefont {A.}~\bibnamefont
  {Abramovici}} \emph {et~al.},\ }\href@noop {} {\bibfield  {journal} {\bibinfo
   {journal} {Science}\ }\textbf {\bibinfo {volume} {256}},\ \bibinfo {pages}
  {325} (\bibinfo {year} {1992})}\BibitemShut {NoStop}%
\bibitem [{\citenamefont {Bradaschia}\ \emph {et~al.}(1990)\citenamefont
  {Bradaschia} \emph {et~al.}}]{virgo}%
  \BibitemOpen
  \bibfield  {author} {\bibinfo {author} {\bibfnamefont {C.}~\bibnamefont
  {Bradaschia}} \emph {et~al.},\ }\href@noop {} {\bibfield  {journal} {\bibinfo
   {journal} {Nucl.\ Instrum.\ Methods Phys.\ Sect A}\ }\textbf {\bibinfo
  {volume} {289}},\ \bibinfo {pages} {518} (\bibinfo {year}
  {1990})}\BibitemShut {NoStop}%
\bibitem [{\citenamefont {Callen}\ and\ \citenamefont {Welton}(1951)}]{fdt}%
  \BibitemOpen
  \bibfield  {author} {\bibinfo {author} {\bibfnamefont {H.~B.}\ \bibnamefont
  {Callen}}\ and\ \bibinfo {author} {\bibfnamefont {T.~A.}\ \bibnamefont
  {Welton}},\ }\href@noop {} {\bibfield  {journal} {\bibinfo  {journal} {Phys.\
  Rev.}\ }\textbf {\bibinfo {volume} {83}},\ \bibinfo {pages} {34} (\bibinfo
  {year} {1951})}\BibitemShut {NoStop}%
\bibitem [{\citenamefont {Uchiyama}\ \emph {et~al.}(1998)\citenamefont
  {Uchiyama} \emph {et~al.}}]{sapphire_cooling}%
  \BibitemOpen
  \bibfield  {author} {\bibinfo {author} {\bibfnamefont {T.}~\bibnamefont
  {Uchiyama}} \emph {et~al.},\ }\href@noop {} {\bibfield  {journal} {\bibinfo
  {journal} {Phys.\ Lett.\ A}\ }\textbf {\bibinfo {volume} {242}},\ \bibinfo
  {pages} {211} (\bibinfo {year} {1998})}\BibitemShut {NoStop}%
\bibitem [{\citenamefont {Uchiyama}\ \emph {et~al.}(1999)\citenamefont
  {Uchiyama} \emph {et~al.}}]{sapphire_subQ}%
  \BibitemOpen
  \bibfield  {author} {\bibinfo {author} {\bibfnamefont {T.}~\bibnamefont
  {Uchiyama}} \emph {et~al.},\ }\href@noop {} {\bibfield  {journal} {\bibinfo
  {journal} {Phys.\ Lett.\ A}\ }\textbf {\bibinfo {volume} {261}},\ \bibinfo
  {pages} {5} (\bibinfo {year} {1999})}\BibitemShut {NoStop}%
\bibitem [{\citenamefont {Uchiyama}\ \emph {et~al.}(2000)\citenamefont
  {Uchiyama} \emph {et~al.}}]{sapphire_fiberQ}%
  \BibitemOpen
  \bibfield  {author} {\bibinfo {author} {\bibfnamefont {T.}~\bibnamefont
  {Uchiyama}} \emph {et~al.},\ }\href@noop {} {\bibfield  {journal} {\bibinfo
  {journal} {Phys.\ Lett.\ A}\ }\textbf {\bibinfo {volume} {273}},\ \bibinfo
  {pages} {310} (\bibinfo {year} {2000})}\BibitemShut {NoStop}%
\bibitem [{\citenamefont {Yamamoto}\ \emph {et~al.}(2006)\citenamefont
  {Yamamoto} \emph {et~al.}}]{coating_Q}%
  \BibitemOpen
  \bibfield  {author} {\bibinfo {author} {\bibfnamefont {K.}~\bibnamefont
  {Yamamoto}} \emph {et~al.},\ }\href@noop {} {\bibfield  {journal} {\bibinfo
  {journal} {Phys.\ Rev.\ D}\ }\textbf {\bibinfo {volume} {74}},\ \bibinfo
  {pages} {022002} (\bibinfo {year} {2006})}\BibitemShut {NoStop}%
\bibitem [{\citenamefont {Tomaru}\ \emph
  {et~al.}(2002{\natexlab{a}})\citenamefont {Tomaru} \emph
  {et~al.}}]{sapphire_thermalCond}%
  \BibitemOpen
  \bibfield  {author} {\bibinfo {author} {\bibfnamefont {T.}~\bibnamefont
  {Tomaru}} \emph {et~al.},\ }\href@noop {} {\bibfield  {journal} {\bibinfo
  {journal} {Phys.\ Lett.\ A}\ }\textbf {\bibinfo {volume} {301}},\ \bibinfo
  {pages} {215} (\bibinfo {year} {2002}{\natexlab{a}})}\BibitemShut {NoStop}%
\bibitem [{\citenamefont {Tomaru}\ \emph {et~al.}(2001)\citenamefont {Tomaru}
  \emph {et~al.}}]{sapphire_optabsorb}%
  \BibitemOpen
  \bibfield  {author} {\bibinfo {author} {\bibfnamefont {T.}~\bibnamefont
  {Tomaru}} \emph {et~al.},\ }\href@noop {} {\bibfield  {journal} {\bibinfo
  {journal} {Phys.\ Lett.\ A}\ }\textbf {\bibinfo {volume} {283}},\ \bibinfo
  {pages} {80} (\bibinfo {year} {2001})}\BibitemShut {NoStop}%
\bibitem [{\citenamefont {Miyoki}\ \emph {et~al.}(2001)\citenamefont {Miyoki}
  \emph {et~al.}}]{cryogenic_contami}%
  \BibitemOpen
  \bibfield  {author} {\bibinfo {author} {\bibfnamefont {S.}~\bibnamefont
  {Miyoki}} \emph {et~al.},\ }\href@noop {} {\bibfield  {journal} {\bibinfo
  {journal} {Cryogenics}\ }\textbf {\bibinfo {volume} {41}},\ \bibinfo {pages}
  {415} (\bibinfo {year} {2001})}\BibitemShut {NoStop}%
\bibitem [{\citenamefont {Ohashi}\ \emph {et~al.}(2003)\citenamefont {Ohashi}
  \emph {et~al.}}]{clik_lism}%
  \BibitemOpen
  \bibfield  {author} {\bibinfo {author} {\bibfnamefont {M.}~\bibnamefont
  {Ohashi}} \emph {et~al.},\ }\href@noop {} {\bibfield  {journal} {\bibinfo
  {journal} {Class.\ Quantum Grav.}\ }\textbf {\bibinfo {volume} {20}},\
  \bibinfo {pages} {S599} (\bibinfo {year} {2003})}\BibitemShut {NoStop}%
\bibitem [{\citenamefont {Tomaru}\ \emph
  {et~al.}(2002{\natexlab{b}})\citenamefont {Tomaru} \emph
  {et~al.}}]{thermal_lens}%
  \BibitemOpen
  \bibfield  {author} {\bibinfo {author} {\bibfnamefont {T.}~\bibnamefont
  {Tomaru}} \emph {et~al.},\ }\href@noop {} {\bibfield  {journal} {\bibinfo
  {journal} {Class.\ Quantum Grav.}\ }\textbf {\bibinfo {volume} {19}},\
  \bibinfo {pages} {2045} (\bibinfo {year} {2002}{\natexlab{b}})}\BibitemShut
  {NoStop}%
\bibitem [{\citenamefont {Agatsuma}\ \emph {et~al.}(2010)\citenamefont
  {Agatsuma} \emph {et~al.}}]{clio}%
  \BibitemOpen
  \bibfield  {author} {\bibinfo {author} {\bibfnamefont {K.}~\bibnamefont
  {Agatsuma}} \emph {et~al.},\ }\href@noop {} {\bibfield  {journal} {\bibinfo
  {journal} {Class.\ Quantum Grav.}\ }\textbf {\bibinfo {volume} {27}},\
  \bibinfo {pages} {084022} (\bibinfo {year} {2010})}\BibitemShut {NoStop}%
\bibitem [{\citenamefont {Yamamoto}\ \emph {et~al.}(2008)\citenamefont
  {Yamamoto} \emph {et~al.}}]{yamamoto_clio}%
  \BibitemOpen
  \bibfield  {author} {\bibinfo {author} {\bibfnamefont {K.}~\bibnamefont
  {Yamamoto}} \emph {et~al.},\ }\href@noop {} {\bibfield  {journal} {\bibinfo
  {journal} {J.\ Phys.: Conf.\ Ser.}\ }\textbf {\bibinfo {volume} {122}},\
  \bibinfo {pages} {012002} (\bibinfo {year} {2008})}\BibitemShut {NoStop}%
\bibitem [{\citenamefont {Uchiyama}\ \emph {et~al.}(2006)\citenamefont
  {Uchiyama} \emph {et~al.}}]{suspension_clio}%
  \BibitemOpen
  \bibfield  {author} {\bibinfo {author} {\bibfnamefont {T.}~\bibnamefont
  {Uchiyama}} \emph {et~al.},\ }\href@noop {} {\bibfield  {journal} {\bibinfo
  {journal} {J.\ Phys.: Conf.\ Ser.}\ }\textbf {\bibinfo {volume} {32}},\
  \bibinfo {pages} {259} (\bibinfo {year} {2006})}\BibitemShut {NoStop}%
\bibitem [{\citenamefont {Ikushima}\ \emph {et~al.}(2008)\citenamefont
  {Ikushima} \emph {et~al.}}]{cryocooler}%
  \BibitemOpen
  \bibfield  {author} {\bibinfo {author} {\bibfnamefont {Y.}~\bibnamefont
  {Ikushima}} \emph {et~al.},\ }\href@noop {} {\bibfield  {journal} {\bibinfo
  {journal} {Cryogenics}\ }\textbf {\bibinfo {volume} {48}},\ \bibinfo {pages}
  {406} (\bibinfo {year} {2008})}\BibitemShut {NoStop}%
\bibitem [{\citenamefont {Braginsky}\ \emph {et~al.}(1999)\citenamefont
  {Braginsky}, \citenamefont {Gorodetsky},\ and\ \citenamefont
  {Vyatchanin}}]{braginsky}%
  \BibitemOpen
  \bibfield  {author} {\bibinfo {author} {\bibfnamefont {V.~B.}\ \bibnamefont
  {Braginsky}}, \bibinfo {author} {\bibfnamefont {M.~L.}\ \bibnamefont
  {Gorodetsky}}, \ and\ \bibinfo {author} {\bibfnamefont {S.~P.}\ \bibnamefont
  {Vyatchanin}},\ }\href@noop {} {\bibfield  {journal} {\bibinfo  {journal}
  {Phys.\ Lett.\ A}\ }\textbf {\bibinfo {volume} {264}},\ \bibinfo {pages} {1}
  (\bibinfo {year} {1999})}\BibitemShut {NoStop}%
\bibitem [{\citenamefont {Cerdonio}\ \emph {et~al.}(2001)\citenamefont
  {Cerdonio}, \citenamefont {Conti}, \citenamefont {Heidmann},\ and\
  \citenamefont {Pinard}}]{cerdonio}%
  \BibitemOpen
  \bibfield  {author} {\bibinfo {author} {\bibfnamefont {M.}~\bibnamefont
  {Cerdonio}}, \bibinfo {author} {\bibfnamefont {L.}~\bibnamefont {Conti}},
  \bibinfo {author} {\bibfnamefont {A.}~\bibnamefont {Heidmann}}, \ and\
  \bibinfo {author} {\bibfnamefont {M.}~\bibnamefont {Pinard}},\ }\href@noop {}
  {\bibfield  {journal} {\bibinfo  {journal} {Phys.\ Rev.\ D}\ }\textbf
  {\bibinfo {volume} {63}},\ \bibinfo {pages} {082003} (\bibinfo {year}
  {2001})}\BibitemShut {NoStop}%
\bibitem [{\citenamefont {Black}\ \emph {et~al.}(2004)\citenamefont {Black},
  \citenamefont {Villar},\ and\ \citenamefont {Libbrecht}}]{black}%
  \BibitemOpen
  \bibfield  {author} {\bibinfo {author} {\bibfnamefont {E.~D.}\ \bibnamefont
  {Black}}, \bibinfo {author} {\bibfnamefont {A.}~\bibnamefont {Villar}}, \
  and\ \bibinfo {author} {\bibfnamefont {K.~G.}\ \bibnamefont {Libbrecht}},\
  }\href@noop {} {\bibfield  {journal} {\bibinfo  {journal} {Phys.\ Rev.\
  Lett.}\ }\textbf {\bibinfo {volume} {93}},\ \bibinfo {pages} {241101}
  (\bibinfo {year} {2004})}\BibitemShut {NoStop}%
\bibitem [{\citenamefont {Nakagawa}\ \emph {et~al.}(2002)\citenamefont
  {Nakagawa}, \citenamefont {Gretarsson}, \citenamefont {Gustafson},\ and\
  \citenamefont {Fejer}}]{nakagawa}%
  \BibitemOpen
  \bibfield  {author} {\bibinfo {author} {\bibfnamefont {N.}~\bibnamefont
  {Nakagawa}}, \bibinfo {author} {\bibfnamefont {A.~M.}\ \bibnamefont
  {Gretarsson}}, \bibinfo {author} {\bibfnamefont {E.~K.}\ \bibnamefont
  {Gustafson}}, \ and\ \bibinfo {author} {\bibfnamefont {M.~M.}\ \bibnamefont
  {Fejer}},\ }\href@noop {} {\bibfield  {journal} {\bibinfo  {journal} {Phys.\
  Rev.\ D}\ }\textbf {\bibinfo {volume} {65}},\ \bibinfo {pages} {102001}
  (\bibinfo {year} {2002})}\BibitemShut {NoStop}%
\bibitem [{\citenamefont {Touloukian}\ \emph {et~al.}(1970)\citenamefont
  {Touloukian}, \citenamefont {Kirby}, \citenamefont {Taylor},\ and\
  \citenamefont {Lee}}]{data_book}%
  \BibitemOpen
  \bibfield  {author} {\bibinfo {author} {\bibfnamefont {Y.~S.}\ \bibnamefont
  {Touloukian}}, \bibinfo {author} {\bibfnamefont {R.~K.}\ \bibnamefont
  {Kirby}}, \bibinfo {author} {\bibfnamefont {E.~R.}\ \bibnamefont {Taylor}}, \
  and\ \bibinfo {author} {\bibfnamefont {T.~Y.~R.}\ \bibnamefont {Lee}},\
  }\href@noop {} {\emph {\bibinfo {title} {Thermophysical Properties of Matter
  - the TPRC Data Series}}}\ (\bibinfo  {publisher} {IFI/Plenum},\ \bibinfo
  {address} {New York},\ \bibinfo {year} {1970})\BibitemShut {NoStop}%
\bibitem [{\citenamefont {Taylor}\ \emph {et~al.}(1996)\citenamefont {Taylor},
  \citenamefont {Notcutt}, \citenamefont {Wong}, \citenamefont {Mann},\ and\
  \citenamefont {Blair}}]{th_expansion_cryo}%
  \BibitemOpen
  \bibfield  {author} {\bibinfo {author} {\bibfnamefont {C.~T.}\ \bibnamefont
  {Taylor}}, \bibinfo {author} {\bibfnamefont {M.}~\bibnamefont {Notcutt}},
  \bibinfo {author} {\bibfnamefont {E.~K.}\ \bibnamefont {Wong}}, \bibinfo
  {author} {\bibfnamefont {A.~G.}\ \bibnamefont {Mann}}, \ and\ \bibinfo
  {author} {\bibfnamefont {D.~G.}\ \bibnamefont {Blair}},\ }\href@noop {}
  {\bibfield  {journal} {\bibinfo  {journal} {Optics Communications}\ }\textbf
  {\bibinfo {volume} {131}},\ \bibinfo {pages} {311} (\bibinfo {year}
  {1996})}\BibitemShut {NoStop}%
\bibitem [{\citenamefont {Saulson}(1994)}]{salson_book}%
  \BibitemOpen
  \bibfield  {author} {\bibinfo {author} {\bibfnamefont {P.~R.}\ \bibnamefont
  {Saulson}},\ }\href@noop {} {\emph {\bibinfo {title} {Fundamentals of
  Interferometric Gravitational Wave Detectors}}}\ (\bibinfo  {publisher}
  {World Scientific},\ \bibinfo {address} {Singapore},\ \bibinfo {year}
  {1994})\BibitemShut {NoStop}%
\bibitem [{\citenamefont {Kuroda}\ \emph {et~al.}(2010)\citenamefont {Kuroda}
  \emph {et~al.}}]{lcgt}%
  \BibitemOpen
  \bibfield  {author} {\bibinfo {author} {\bibfnamefont {K.}~\bibnamefont
  {Kuroda}} \emph {et~al.},\ }\href@noop {} {\bibfield  {journal} {\bibinfo
  {journal} {Class.\ Quantum Grav.}\ }\textbf {\bibinfo {volume} {27}},\
  \bibinfo {pages} {084004} (\bibinfo {year} {2010})}\BibitemShut {NoStop}%
\bibitem [{\citenamefont {Punturo}\ \emph {et~al.}(2010)\citenamefont {Punturo}
  \emph {et~al.}}]{et}%
  \BibitemOpen
  \bibfield  {author} {\bibinfo {author} {\bibfnamefont {M.}~\bibnamefont
  {Punturo}} \emph {et~al.},\ }\href@noop {} {\bibfield  {journal} {\bibinfo
  {journal} {Class.\ Quantum Grav.}\ }\textbf {\bibinfo {volume} {27}},\
  \bibinfo {pages} {084007} (\bibinfo {year} {2010})}\BibitemShut {NoStop}%
\end{thebibliography}%

\end{document}